\documentclass[aps,prb,twocolumn,showpacs]{revtex4}
\usepackage{graphicx} 
\begin{document}

\title{Slow dynamics of a confined supercooled binary mixture II: 
$Q$ space analysis}

\author{P.~Gallo$^\dagger$\footnote[1]{Author to whom correspondence 
should be addressed; e-mail: gallop@fis.uniroma3.it
}, R.~Pellarin $^\dagger$ and M.~Rovere$^\dagger$,}
\affiliation{$\dagger$ Dipartimento di Fisica, 
Universit\`a ``Roma Tre'', \\ 
Istituto Nazionale per la Fisica della Materia,\\
Unit\`a di Ricerca Roma Tre\\
and Democritos National Simulation Center \\
Via della Vasca Navale 84, 00146 Roma, Italy.}

\begin{abstract}
We report the analysis in the wavevector space of the density correlator
of a Lennard Jones binary mixture confined in a disordered 
matrix of soft spheres upon supercooling. 
In spite of the strong confining medium 
the behavior of the mixture is consistent with the Mode Coupling Theory 
predictions for bulk supercooled liquids.
The relaxation times extracted from the fit 
of the density correlator to the
stretched exponential function 
follow a unique power law behavior as a function of 
wavevector and temperature. The von Schweidler scaling properties are
valid for an extended wavevector range around
the peak of the structure factor. 
The parameters extracted in the present work are compared with
the bulk values obtained in literature. 
\end{abstract}


\maketitle

\section{Introduction}

Modifications of vitrification properties of liquids under
confinement are of interest for relevant technological 
and biophysical problems \cite{pablo,grenoble}.
While experiments show that the glass transition 
is not usually suppressed when the glass former is confined,
changes in the temperature of the transition 
and in dynamical properties of liquids upon supercooling
are induced by both the geometrical confinement and
the interaction with the substrate~\cite{grenoble}. 

In the study of the glass transition phenomenology  
the mode coupling theory (MCT) of
the evolution of glassy dynamics plays a fundamental role.
MCT is able 
to describe the slowing down of supercooled liquids in a region of mild 
supercooling on approaching a temperature $T_C$ \cite{gotze1,gotze2,gotze3}.
This temperature marks a
crossover from a region where dynamics is mastered by the transient
caging of nearest neighbor to a lower temperature region where 
hopping processes dominate. In the ideal version of the theory,
where hopping is not included, 
the non linear feedback mechanisms in the microscopic dynamics of the
particles become so strong that they lead to the structural arrest of the
system. Close to $T_C$ MCT asymptotic solutions show analytic behaviors
of the density correlator from which parameters of the theory
can be extracted. These behaviors have been effectively found
in experiments and simulations on supercooled liquids in regions
close to $T_C$ where hopping was negligible \cite{gotze3}.

The role of the cooperative dynamics of the particles in the
glass transition is still under discussion and there is an increasing
activity on the definition and observation
of dynamical heterogeneities~\cite{adam-gibbs}.
Cooperativity in the dynamical behavior of the
particles increases as the system is approaching the glass transition,
suggesting the idea of the existence of
a dynamic correlation length~\cite{donati1,donati2,doliwa1,doliwa2,berthier}. 

Molecular dynamics studies of model liquids in restricted
geometries intended to assess the applicability of MCT 
can give an important contribution to the 
characterization of vitrification 
processes in confinement~\cite{gallo1,gallo2,baschnagel,kobconf1,kobconf2}.  

Liquids confined in network
of interconnected pores with a large value of porosity, as Silica Xerogels, 
can be appropriately studied with models where the confining
solid is built as a disordered array of 
frozen microspheres~\cite{rosinberg,monson,page}. 
We consider here one of such system. A liquid Lennard Jones binary 
mixture (LJBM) 
composed of $80\%$ of particles A and $20 \%$ of smaller particles B
embedded in an off lattice matrix of soft spheres.  
We performed Molecular Dynamics (MD) simulations 
of the confined LJBM upon supercooling in order to test the 
predictions of MCT. 

The bulk phase of this LJBM behaves 
{\it \`a la Mode Coupling}~\cite{kob,kob1,kob2}. 
A preliminary analysis on our system \cite{euro} showed evidences that 
predictions of MCT hold also for the confined LJBM.
A detailed study on the dynamics carried out in the direct 
space has shown
two important differences with respect to the 
bulk upon supercooling~\cite{noi}.
The smaller B particles tend to avoid the soft sphere interfaces 
on lowering temperature and correspondingly their
diffusion coefficient becomes lower than that of the A particles,
at variance with the bulk where there is no such an inversion.
Hopping processes are markedly present for both A and
B particles already above $T_C$ while 
in the bulk system only B particles show consistent hopping.

As a possible consequence of these differences, we observed 
that the range of validity of the power law fit to the diffusion
coefficient as predicted by MCT 
in the long time region, late $\alpha$ relaxation region,
appears much more limited than in the bulk. 

A more refined assessment of the MCT validity can be obtained 
by a quantitative comparison between the asymptotic
predictions of the theory and the results of our
computer simulation in the $(Q,t)$ space,
where the density correlators of a liquid
testing MCT display clear signatures of 
the approach to $T_C$.
In this space important scaling relations of the
theory can be verified. For these reasons we carry on
here an analysis of the behavior of the
self intermediate scattering functions, $F_S(Q,t)$.

In the next section we recall the predictions of the idealized version of MCT
connected to our analysis. 
In Sec. III we describe our model system and
Molecular Dynamics details. In Sec. IV we perform a test of MCT properties
on our system. In particular we analyze the static structure factor
and the self intermediate scattering function. For this latter quantity
we perform numerical fit to MCT analytical laws and extract
parameters of the theory.
Last section is devoted to conclusions.

\section{MCT scenario for the evolution of the structural
relaxation}

The MCT for the evolution of glassy dynamics developed originally
to deal with the cage effect provides a model for an ideal
liquid-glass transition. The evolution of the structural
relaxation is described in terms of the asymptotic behavior
of density correlators near singularities 
which take place close to a cross-over temperature $T_C$. 

For a density correlator or any other
correlator that couples with density fluctuations, $\phi(Q,t)$, 
MCT predicts upon supercooling a diversification
of relaxation times related to the nearest neighbor caging.
In the short time regime  
the correlator show a fast relaxation associated with  
the ballistic motion of the particles. 
The high temperature correlator relaxes then 
exponentially for longer times.
On lowering temperature for intermediate times a 
shoulder appears in the relaxation law. 
As the temperature is further lowered the shoulder enhances and becomes 
a plateau and the correlation function clearly   
shows a two step relaxation. The presence of the 
plateau is related to the trapping of
the particle in the cage formed by its nearest neighbors.
The time interval in which the correlation function is close-to 
and above the plateau
is called the $\beta$ relaxation region. The region that 
starts when the correlation function leaves the plateau
is called the $\alpha$ relaxation region.

The long time limit of $\phi(Q,t)$ is the non ergodicity parameter
$f_Q(T)$. This parameter exhibits a singularity 
at a temperature $T_C$ where the system undergoes an ideal transition
from an ergodic to a non ergodic behavior. $f_Q(T)$ jumps
discontinuously from $0$ above $T_C$ to $f_Q^c$ at $T_C$.
On approaching  $T_C$ from below
\begin{equation}
f_Q-f_Q^c \sim \sqrt{\vert \epsilon \vert} \label{eq:fq}
\end{equation}
where $\epsilon= (T-T_C)/T_C$ is the small parameter of 
the theory and $f_Q^c$ is called the critical non ergodicity parameter.
On approaching $T_C$ from above an effective $f_Q$ can be identified
with the height of the plateau of the correlator. In this range
of temperatures the values of $f_Q$ 
are very close to $f_Q^c$ and show a very mild temperature
dependence
\begin{equation}
f_Q-f_Q^c \sim O (\epsilon ) \label{eq:fq1}
\end{equation}
In the asymptotic limit where $\epsilon << 1$ MCT predicts
a factorization 
\begin{equation}
\phi(Q,t)-f_Q^c = h_Q G(t) \label{eq:gt}
\end{equation}
where $h_Q$ is a positive factor which does not depend on time.
The so called $\beta$ correlator, G(t), obeys the scaling
relation 
\begin{equation}
G(t) \sim \sqrt{\vert \epsilon \vert} g(t/t_\epsilon) \label{eq:gt1} 
\end{equation}
where the time scale $t_\epsilon$ has been introduced
\begin{equation}
t_\epsilon \sim \frac{t_0}{{\vert \epsilon \vert}^{1/2a}}.
\end{equation}
Here $t_0$ is a microscopic characteristic time of the system and
the exponent $a$ is called the critical exponent. 
In the time scale which corresponds to the decay
of the correlator towards the plateau, $t \le t_\epsilon$
\begin{equation}
g \left( t/t_\epsilon \right)= \left( t_\epsilon/t \right)^a  \label{eq:gta} 
\end{equation}

For the correlator decay
of the liquid from the plateau  
\begin{equation}
g \left( t/t_\epsilon \right) \sim
 \left( t/t_\epsilon \right)^b  \label{eq:gtb}
\end{equation}
which is valid for times larger than the time scale of the $\beta$
relaxation and much smaller than the $\alpha$-relaxation time $\tau$,
 $t_\epsilon \le t << \tau$.
The exponent $b$ is known as the von Schweidler exponent.
The critical and the von Schweidler exponents are related by
the formula
\begin{equation}
\lambda = \frac{[ \Gamma \left( 1-a \right)]^2}{\Gamma \left( 1-2a \right)}
= \frac{[\Gamma \left( 1+b \right)]^2}{\Gamma \left( 1+2b \right)}
\label{eq:lambda}
\end{equation}
where $\lambda$ is called the exponent parameter and $\Gamma(x)$ 
is the Euler Gamma function.

In the late part 
of the $\alpha$-relaxation regime the shape of the
correlator of a liquid approaching the
MCT crossover temperature~\cite{gotze2} can be represented
by means of the Kohlrausch-Williams-Watts (KWW) analytical function
\begin{equation}
\phi(Q,t)=f_Q \exp(-(t/\tau)^\beta) \label{eq:kww}
\end{equation} 
where $\beta$ is the Kohlrausch exponent. 

The $\alpha$-relaxation time $\tau$ is expected from MCT to follow the 
power law 
\begin{equation}
\tau^{-1} \sim (T-T_C)^{\gamma} \label{eq:tau}
\end{equation}
Both the exponent $\gamma$ and the crossover temperature $T_C$
should coincide with the ones appearing 
in the temperature dependence of
the diffusion coefficient $D \sim (T-T_C)^{\gamma}$.

An important prediction of MCT is the validity of 
the time-temperature superposition principle. It states that in the low 
temperature regime, approaching from above the MCT crossover temperature, 
it is possible to time-rescale 
the correlators evaluated at different T
into a single master curve:
\begin{equation}
\phi(t)=\tilde\phi(t/\tau(T)) \label{eq:master}
\end{equation}
$\tilde\phi$ is the master function which in the
late $\alpha$-relaxation limit becomes the KWW function~(\ref{eq:kww}).

In the late $\beta$, early $\alpha$
relaxation regime Eq.~\ref{eq:gt} can be written as 
\begin{equation}
\phi(Q,t)=f_Q^c-h_Q(t/\tau)^b \label{eq:vsclaw}
\end{equation}
This equation is called the von Schweidler law and $\tau$
indicates a characteristic time of the system.

The exponent $\gamma$ in Eq.~\ref{eq:tau} is related
to the exponents $a$ and $b$ by the equation
\begin{equation}
\gamma=\frac{1}{2a}+\frac{1}{2b} \label{eq:gamma}
\end{equation}

While both $\tau$ and the exponent $\beta$ are
$Q$ dependent, the exponent $b$ should not
depend on $Q$. These two exponents are related: $\beta>b$ and
$\beta(q \rightarrow \infty) = b$~\cite{fucs}.

\section{Model and Molecular Dynamics}

The model studied consists in a 
rigid off lattice matrix of 16 soft spheres 
where  a liquid $A_{80}B_{20}$
Lennard Jones binary mixture (LJBM) \cite{kob1,kob2} 
is embedded.
The mixture is composed
of 1000 particles.
In the following LJ units will be used.
The parameters of the LJBM are 
$\epsilon_{AA}=1$, $\sigma_{AA}=1$, $\epsilon_{BB}=0.5$,
$\sigma_{BB}=0.88$, $\epsilon_{AB}=1.5$, and $\sigma_{AB}=0.8$.
The parameters of the soft sphere potential are
$\epsilon_{SA}=0.32$, $\sigma_{SA}=3$, 
$\epsilon_{SB}=0.22$, $\sigma_{SB}=2.94$.
The box length is L=12.6. 
We performed Molecular dynamics simulations in the NVE ensemble.
We studied a set of temperatures ranging from T=5.0 down to T=0.37. 
The timestep for the lowest temperature is 0.02 and the largest 
production run consists in $14$x$10^6$ timesteps.
For all
temperatures the time range investigated by MD was sufficient for the
correlation functions to decay to zero in the long time 
limit. Further MD simulation details are reported in ref.s~\cite{noi,euro}.
We did not find any
qualitative differences in the behavior of the thermodynamic quantities
with respect to the bulk upon supercooling~\cite{noi}.

\begin{figure}
\includegraphics[width=8cm]{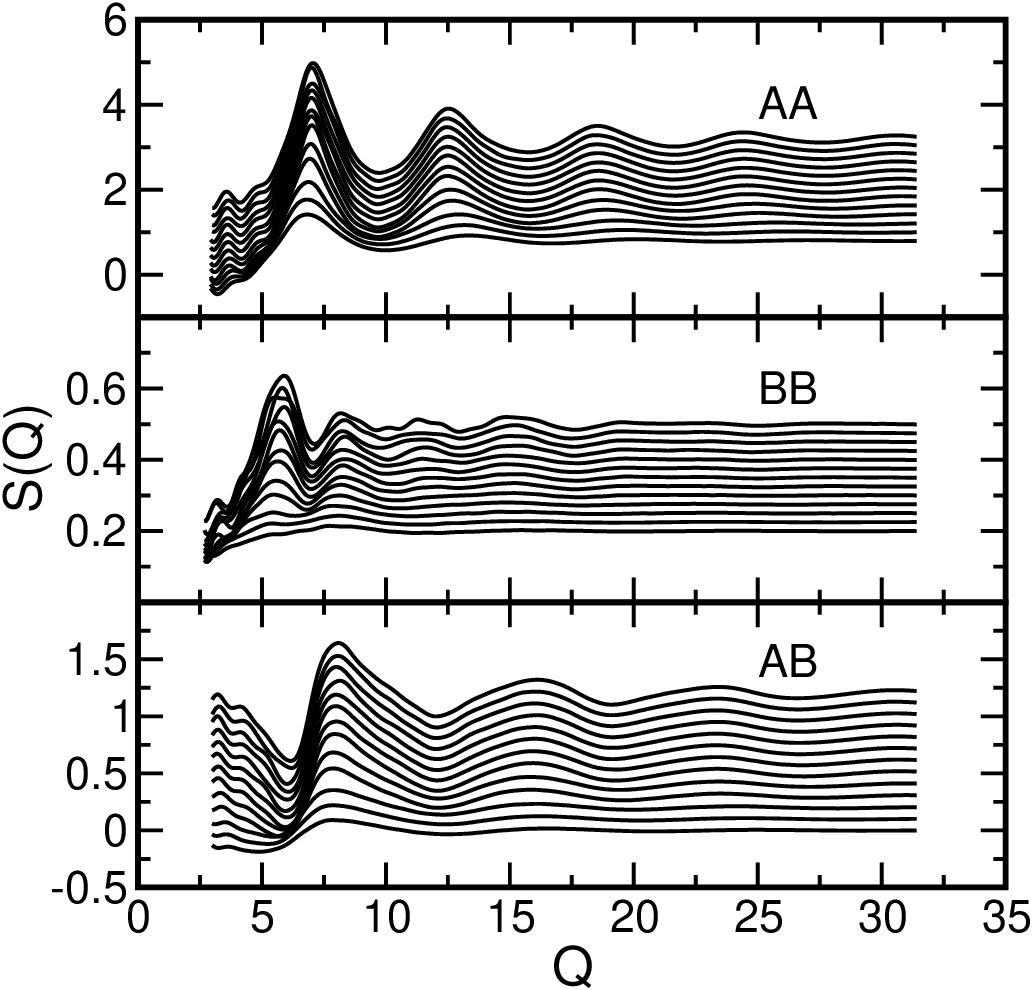}
\caption{Static structure factors. 
The temperatures corresponding to 
different curves are $5.0$, $2.0$, $0.8$, 
$0.58$, $0.538$, $0.48$, $0.465$, $0.43$, $0.41$, $0.39$, $0.37$.
For clarity we shifted 
up the lower temperature plots. For AA the shift is 0.2, 
for BB is 0.025 and for AB is 0.1.}
\protect\label{fig:1}
\end{figure} 

\begin{figure}
\includegraphics [width=8cm]{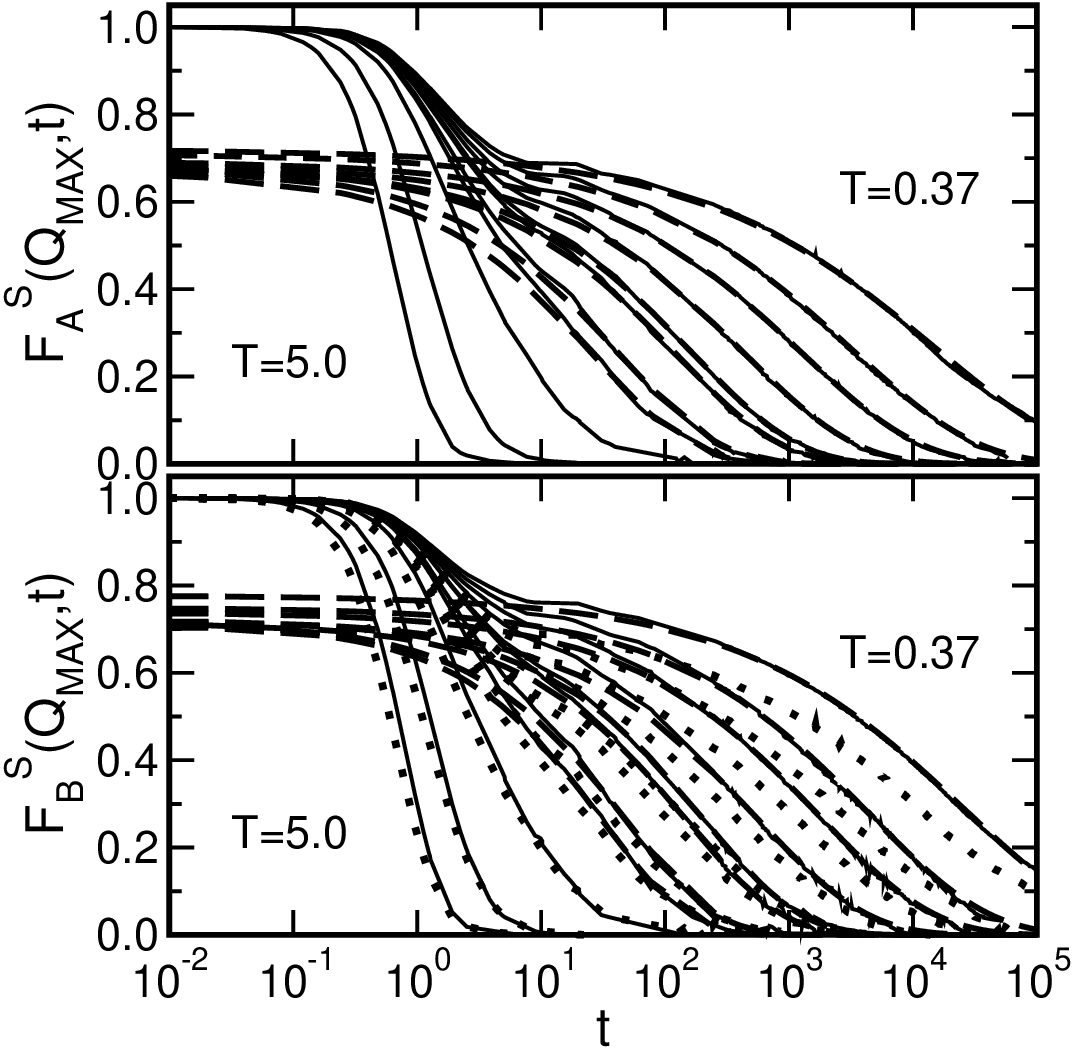}
\caption{Intermediate scattering function
for A particles (upper panel) and for B particles (lower panel)
for all the investigated temperatures. The 
continuous lines are calculated from the MD trajectories
at the peaks of the structure factor
$Q_{MAX}^{A}=7.0$ and $Q_{MAX}^{B}=5.8$. 
The long dashed lines are the fit to the KWW
stretched exponential function, see Eq.\protect\ref{eq:kww}. 
In the lower graph the dotted lines are the A particles 
data, reported for comparison.}
\protect\label{fig:2}
\end{figure} 

\section{Analysis of the dynamical behavior in $(Q,t)$ space}

\subsection{Static structure factor}

In the framework of MCT the static structure factor
is the input information for MCT equations \cite{gotze1,gotze2}.
This function is expected to depend smoothly on wavevectors and 
control parameters and it is also expected to show no singularities 
upon supercooling. In Fig.~\ref{fig:1} we report the structure factors
$S(Q)$ of the confined mixture for the temperatures investigated. 
They behave  consistently with MCT as
the functions smoothly vary upon supercooling implying no divergence
of correlation length. No signature of phase separation 
of the mixture are evident in confinement
as we already deduced from the analysis of the $g_{AB}(r)$~\cite{noi}.
We observe that there are no
relevant discrepancies in the confined $S_{ij}(Q)$ 
with respect to the bulk~\cite{kob2}
neither in the shape nor in the peak positions. 
For $Q$ values around the first structure factor peak
MCT features are best evident in the density correlators. 
The peak positions in our confined LJBM show a very mild 
temperature dependence, especially for low temperatures. 
We will therefore consider them constant. 
The values $Q^{A}_{MAX}=7.06 $
$Q^{B}_{MAX}=5.90 $ will be used
to test MCT as a function of temperature in the following. 
In the case of the bulk the values used 
were respectively $Q^{AA}_{MAX}=7.25$ and $Q^{BB}_{MAX}=5.75$.

\subsection{Self intermediate scattering function}

We consider now the self intermediate scattering function (SISF)
defined as
\begin{equation}
F_S(Q,t)=\frac{1}{N} <\sum_i e^{-i\vec Q\cdot\vec R_i(0)}e^{i\vec Q
\cdot\vec R_i(t)}>
\end{equation}
This correlator can be used to
test the predictions of MCT close to $T_C$ as described 
in Sec. II.  
In Fig.~\ref{fig:2} the SISF of our confined system
is plotted for all considered temperatures and for both particle types
at the peak of the structure factors.
At high temperatures the correlators relax  
exponentially for longer times.
On lowering temperature we observe the onset of the plateau
related to the two step relaxation typical of a supercooled liquid.  

An onset temperature for the plateau of
$T\approx 0.6$  can be identified for our system. 
The corresponding temperature for the bulk is $T\approx 1.0$.
Analogous to bulk supercooled liquids
for the lowest temperature we observe that correlators stretch several 
decades in time. We also see that A and B particles relax in a similar fashion
but larger A particles relax faster at this wavevectors, especially  
for lower temperatures. This counterintuitive phenomenon
that happens only in confinement is possibly connected to 
packing constraints.
We have in fact detected that on lowering temperature smaller B 
particles tend to avoid
the soft sphere interfaces and preferentially move in the center of the 
interstices. Correspondingly 
diffusion coefficients of A and B particles
extracted from the mean square displacement 
(MSD) show an inversion as supercooling progresses~\cite{noi}.

\begin{figure}
\includegraphics[width=8cm]{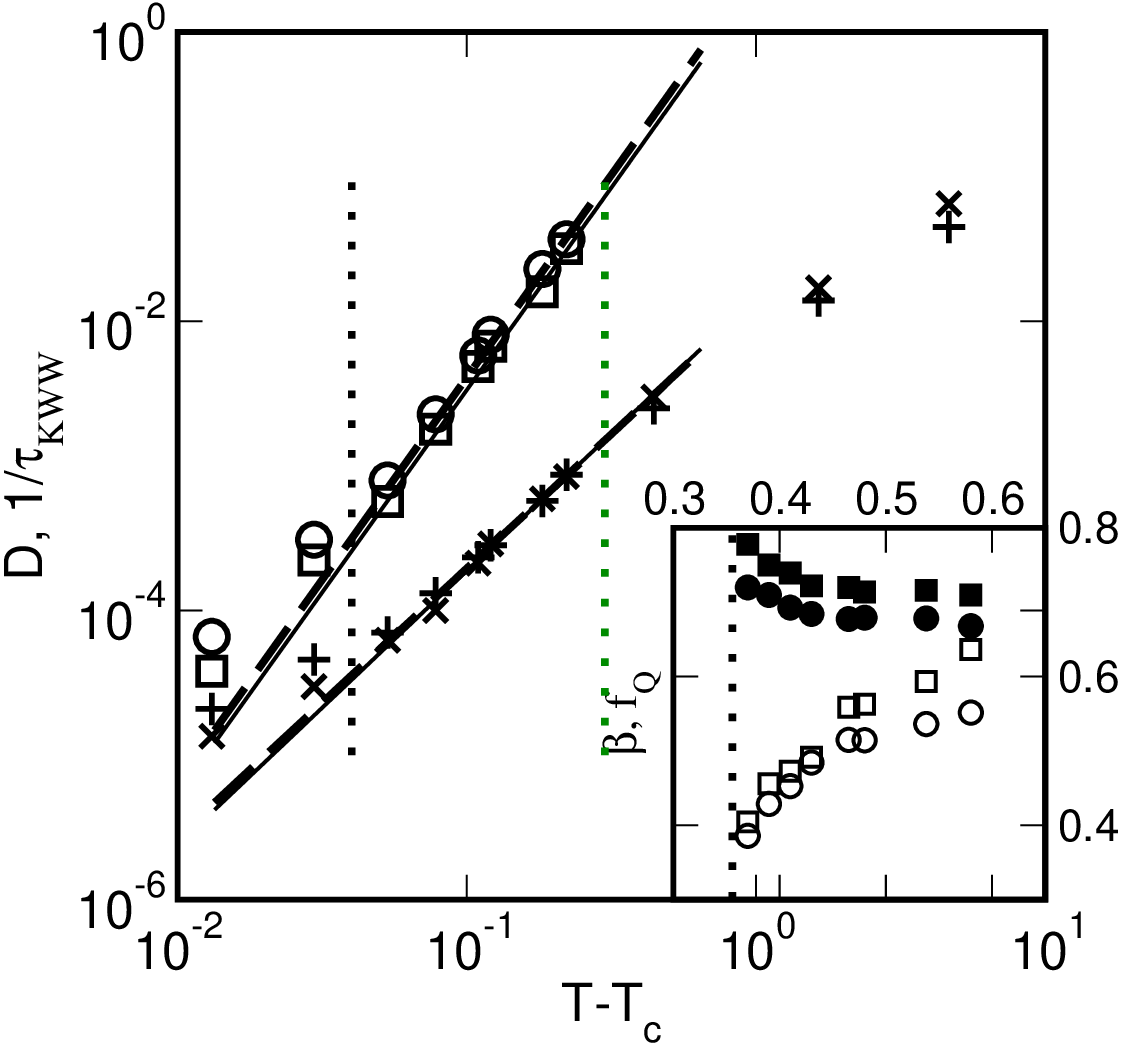}
\caption{In the main frame, values of the relaxation time $\tau$
obtained from the fit to the KWW
of the SISF (see Fig.\protect\ref{fig:2}) for A (empty circles)
and B (empty squares) particles. 
In the same graph values of the diffusion coefficients 
$D$ obtained from the MSD\protect\cite{noi} 
for A (+) and B (X) particles.
The power law fits are also plotted.
These fits are performed on the temperature 
window $0.41 \le T \le 0.58$, limited by vertical dotted lines. 
Dashed lines are the fit for A and continuous lines are
the fit for B particles. 
Parameters obtained from the fit are reported in Tab.\protect\ref{tab:1}.
In the inset $f_Q$ for A (filled circles)
and B (filled squares) particles and stretching exponent $\beta$ 
for A (empty circles) and B (empty squares) particles are reported.}
\protect\label{fig:4}
\end{figure} 

\begin{figure}
\includegraphics[width=8cm]{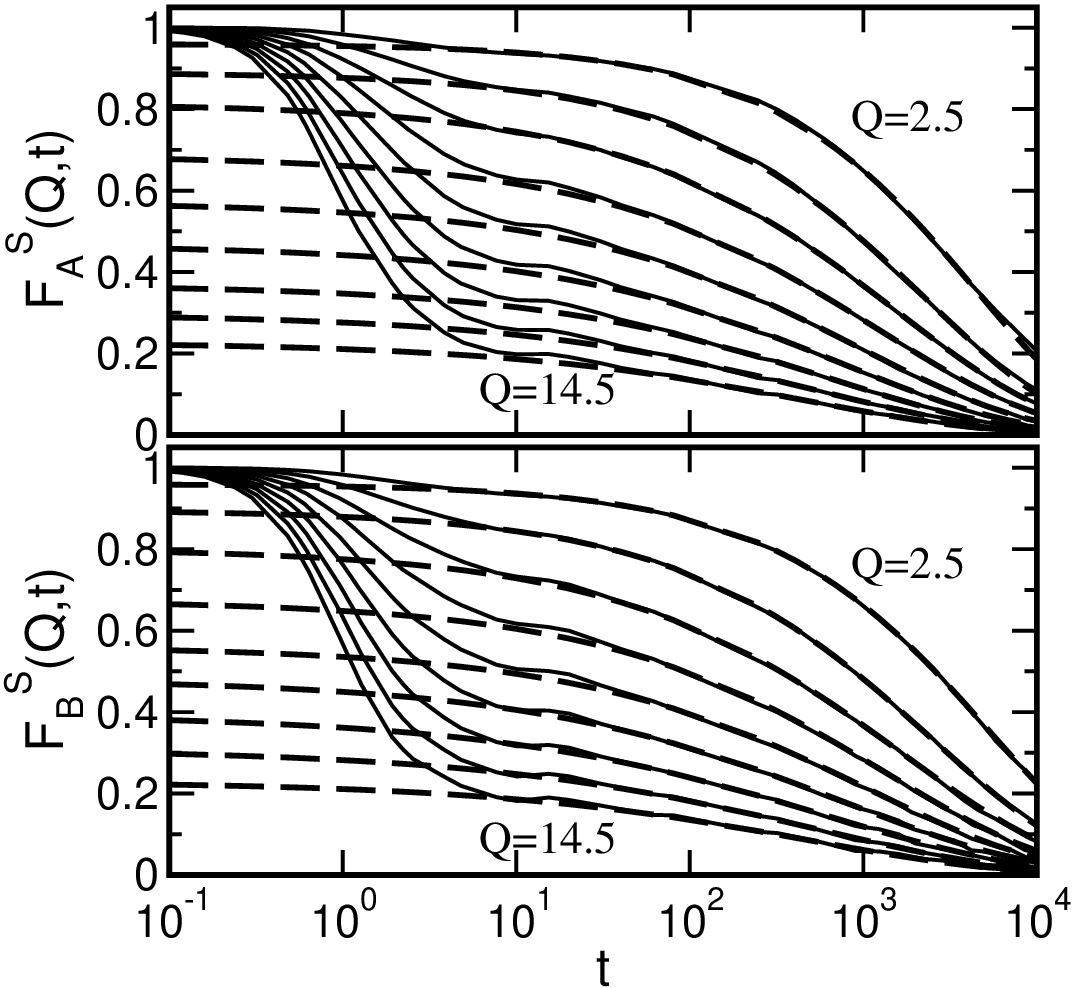}
\caption{Wave vectors analysis at $T=0.41$. 
The range for both
particles is $2.5 \le Q \le 14.5$, with a step of 
$\delta Q = 1.5 $. The KWW fit (long dashed lines) 
is also shown.}
\protect\label{fig:6}
\end{figure} 

\begin{figure}
\includegraphics[width=8cm]{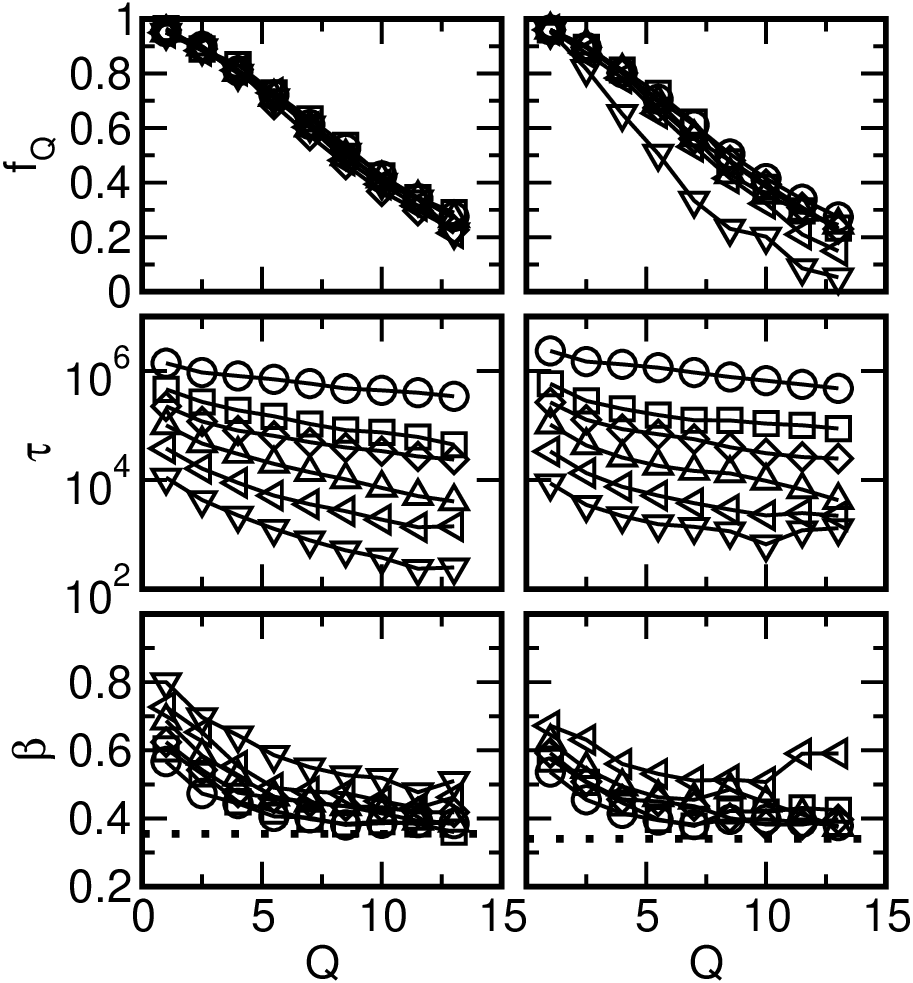}
\caption{In the graphs are shown the KWW parameters $f_Q$, $\tau$, 
$\beta$ upon varying the wave vector $Q$, for A (panels on the left)
and B particles (panels on the right).
The various symbols refer to different temperatures: circle 0.37, square 0.39,
diamond 0.41, triangle up 0.43, triangle left 0.48, triangle down 0.58.
The dotted line of the bottom panels refers to the
values extracted from the von Schweidler fit $b_A=0.355$ and 
$b_B=0.350$.}
\protect\label{fig:7}
\end{figure} 


\subsection{Kohlrausch-Williams-Watts analysis of the 
self intermediate scattering function}

In Fig.~\ref{fig:2} together with the SISF we show the fit to the KWW
function, as defined in Eq.~\ref{eq:kww},
performed for temperatures $T<0.8$ in the $\alpha$ relaxation region
for both A and B particles. 
A remarkable agreement with the data is found.
From the fit the KWW relaxation time $\tau$, the stretching exponent
$\beta$ and the effective non ergodicity
parameter $f_Q$ are extracted.

MCT predicts both $D$ and $\tau^{-1}$ to follow 
the same power law, see Eq.\ref{eq:tau}. 
The relaxation times $\tau$ are shown
in the main frame of Fig.~\ref{fig:4}.
In the same picture we show the diffusion coefficients $D$ 
previously reported and obtained from the
slope of the MSD \cite{noi}. 
We also display in the figure the fit to the MCT predicted 
behavior. Values of $T_C$ and $\gamma$ obtained from the fit are 
reported in Tab.\ref{tab:1}. We observe that both in bulk
and confined system the crossover temperature $T_C$ extracted
from $D$ and $\tau$ and from A and B particles coincide,
$T_C\simeq 0.36$
A substantial reduction of $T_C$ is observed in 
confinement. The bulk value reported in literature is
$T_C\simeq 0.43$.

The exponents $\gamma$ which determine the slope
of the fit are the same for A and B particles
both for $\tau$ and $D$. This agrees with the
prediction of MCT that these exponents should be universal
and do not depend on the species of the particles
involved. A discrepancy with respect to MCT
prediction is represented by the different value of
$\gamma$ between $D$ and $\tau$. As we can see from the
table the same discrepancy affects the mixture also in
the bulk phase and it may be due to long time activated 
processes causing a breakdown of the predictions of the theory only
for quantities calculated at extremely  long times~\cite{noi}. 
In fact in confinement, where hopping is consistent for both A and 
B particles, a reduction of circa $35\%$ is observed for $\gamma$ 
in going from $\tau$ to $D$. In the bulk phase, where hopping is 
more consistent for B particles, the reduction observed is $30\%$ for
B and $20\%$ for A particles.

A substantial shrinkage of the range of validity of
the theory is found in confinement.
In Fig.~\ref{fig:4} we in fact see that the MCT predicted  power law 
behavior is verified for $\tau$ in the temperature window 
$0.58\le T \le 0.41$, the same found for the $D$ coefficients~\cite{noi},
which corresponds to 
$0.153<\epsilon<0.631$ while
in the bulk $0.07<\epsilon<1.30$~\cite{kob}.

The values of the parameters $f_Q$ and $\beta$ extracted from the
fit are  plotted as a function of temperature in 
the inset of Fig.~\ref{fig:4}.
We note that the effective non ergodicity parameter $f_Q$ only 
slightly depends on temperature
in agreement with MCT  that predicts a consistent temperature dependence
of this factor only below $T_C$.
The stretching parameter $\beta$ decreases upon lowering 
the temperature as also expected, but we 
observe no flattening of this exponent
on approaching $T_C$ as found for bulk
systems. Besides, correlators appear much more stretched
in confinement. In the bulk $\beta>0.78$ while here the lowest values 
are slightly below $0.4$.

We report in Fig.~\ref{fig:6} the temporal evolution 
of the correlators for different wavevectors
$Q$ ranging from $2.5$ to $14.5$ for the temperature $T=0.41$.
In the same picture we also show the fit to the KWW function. 
A remarkable agreement is found.
The height of the plateau strongly 
depends on $Q$. The same analysis was performed for 
$T=0.58, 0.48, 0.43, 0.39$ (not shown). 

The values of the parameters of the fit
are reported for all the above temperatures
in Fig.~\ref{fig:7}. All the parameters show a 
monotonic behavior in $Q$. In particular at high temperature  
the relaxation time $\tau$ spans two decades. It is worth noting 
that the $\beta$ exponent is converging, for $T \rightarrow T_C$ and 
for $Q \rightarrow \infty$ to the value of $b$ for both types of 
particles as expected~\cite{fucs}. 
Finally we note that the prefactor $f_Q$ 
is slightly $T$ dependent for all $Q$.

\begin{figure}
\includegraphics[width=7cm]{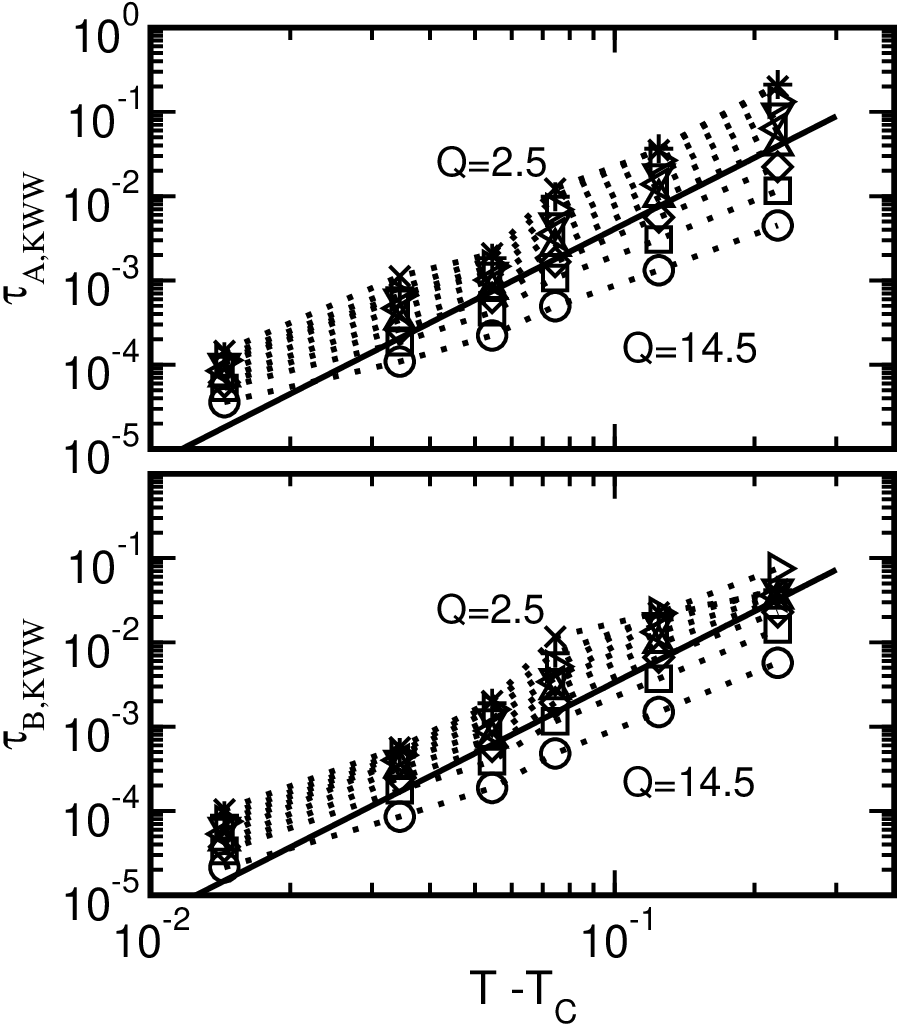}
\caption{Relaxation times $\tau$ (symbols) vs. shifted temperature $T-T_C$.
The power law fit for $Q=Q_{MAX}$ is also plotted. 
Various symbols represent different $Q$. The slope of
the overlaid straight line corresponds to $\gamma =2.9$.}
\protect\label{fig:7a}
\end{figure}

In Fig.~\ref{fig:7a} the $\tau$ values 
extracted from the fit are reported as a function of 
the shifted temperature $T-T_C$. 
In the temperature range where the 
power law behavior is effective, that for our system is 
$0.41 \le T \le 0.58 $, $\gamma$ is expected to be independent on $Q$. 
In Fig.~\ref{fig:7a} we effectively observe that 
the curves are parallel to each other and parallel to
the straight line overlayed that corresponds to the
value of $\gamma=2.9$ extracted from the fit performed
at the peak of the structure factor (see Tab.\ref{tab:1}).

\begin{figure}
\includegraphics[width=8cm]{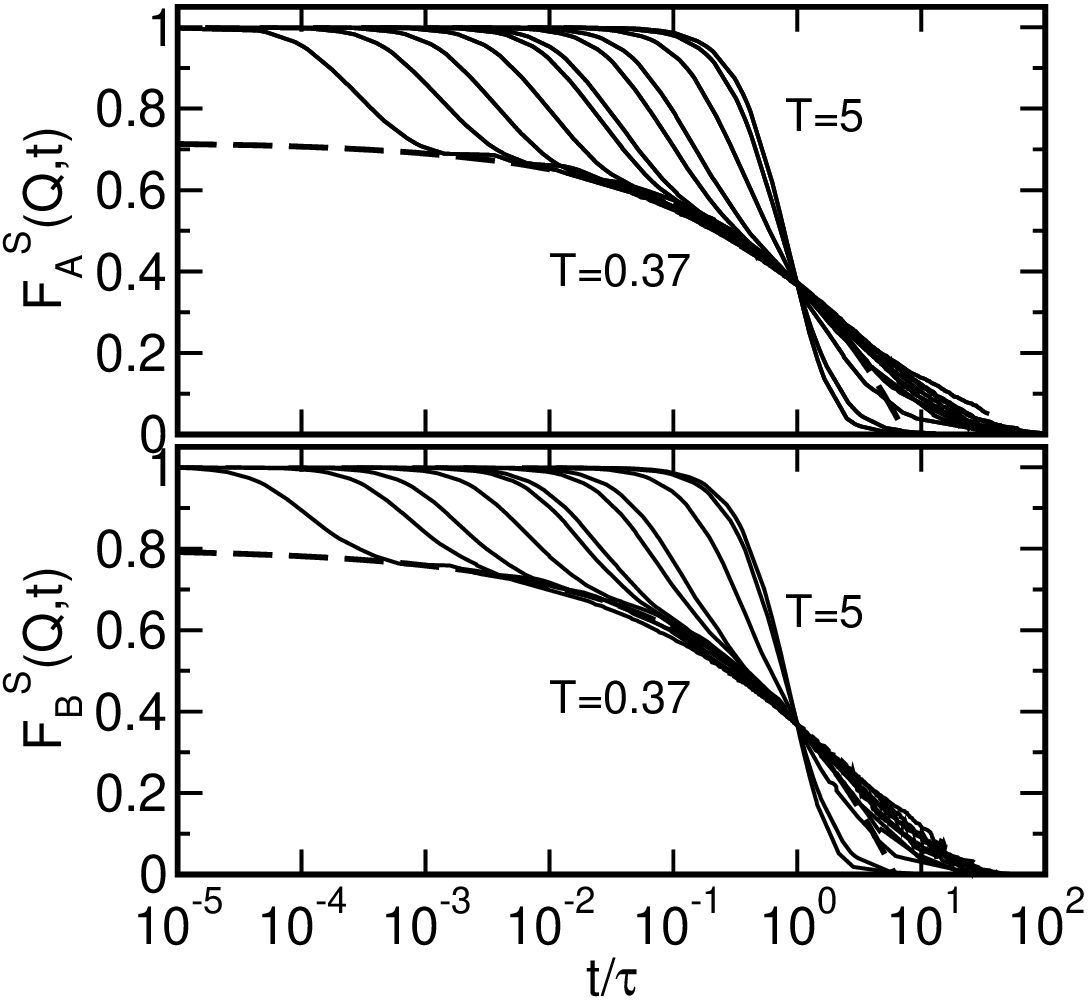}
\caption{Intermediate scattering functions of 
Fig.\protect\ref{fig:2} 
rescaled with respect to the corresponding 
KWW relaxation time for both A (top panel)
and B (bottom panel) particles. The dashed curves 
represent the von Schweidler fit.}
\protect\label{fig:5}
\end{figure} 
\begin{figure}
\includegraphics[width=8cm]{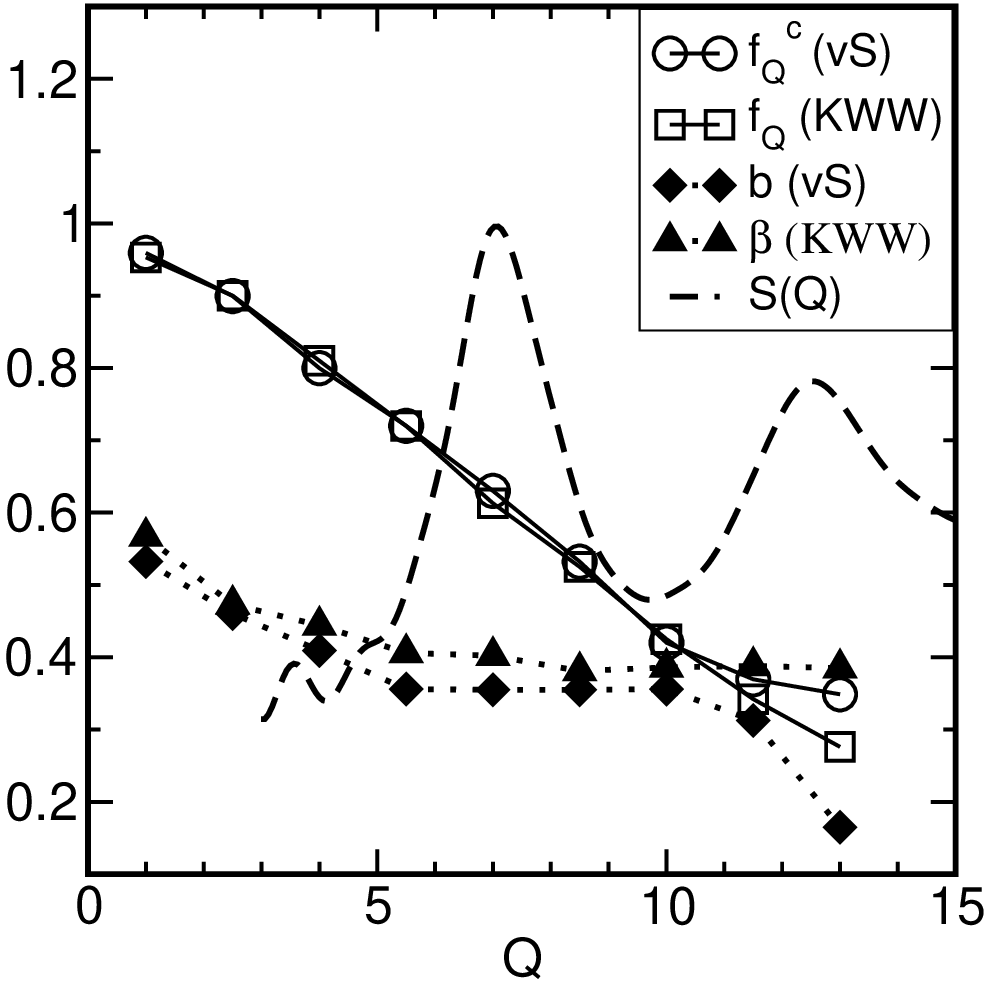}
\caption{Values of the parameters of the von Schweidler fit 
(black diamond, white circles) and for comparison the parameters 
of the KWW (black triangles, white squares) upon varying $Q$ for A particles. 
The static structure factor is represented by the dotted line
at $T=0.37$.}
\protect\label{fig:8}
\end{figure} 

\subsection{von Schweidler test for the self intermediate scattering
function}

According to Eq.~\ref{eq:master} the time 
dependence of the correlation functions can be
rescaled by the relaxation time $\tau$ extracted 
from the fit to the KWW. 
The curves for the lowest temperatures are expected to collapse onto 
a mastercurve in the late $\beta$, early $\alpha$ relaxation region.
In Fig.~\ref{fig:5} we report for both particle types 
the master plot of the SISF of Fig.~\ref{fig:2}. 
In the same figure we show the fit to the von Schweidler 
functional form of~Eq.~\ref{eq:vsclaw}.

The values of the parameters obtained for A and B particles 
from the fit are:
$$f_{Q,A}^c=0.72, \qquad h_{Q,A}=0.35, \qquad b_A=0.355$$
$$f_{Q,B}^c=0.79, \qquad h_{Q,B}=0.44, \qquad b_B=0.350$$
In the bulk the values of the exponent $b$ are respectively $0.51$ for
A and $0.46$ for B particles while we observe that in confinement 
$b$ values are lower and almost the same. This might indicate 
that hopping is influencing also the earlier $\alpha$ relaxation
region.

An important prediction of MCT  is that the von Schweidler $b$ value  
should be independent of the wave vector $Q$. 
This test was performed in the bulk \cite{kob2}, where the 
authors stressed a small dependence of $b$ between $Q_{MAX}$ and 
$Q_{MIN}$, where $Q_{MIN}$
is the position of the first minimum after the peak 
of the static structure factor. 
In Fig.~\ref{fig:8} the values for $f_Q^c$ and $b$ are 
presented as a function of $Q$ for A particles. The values for $f_Q$ 
and $\beta$ for the KWW and the $S(Q)$ are also shown. The $f_Q$ 
factors are in perfect agreement with the Gaussian behavior 
predicted by the MCT and found in the bulk simulations. We can 
therefore consider the $f_Q$ KWW factors as a good approximation 
of the critical non-ergodicity parameter, $f_Q^c$, at such low 
temperature. The $b$ parameter decreases monotonically, except 
between the first $S(Q)$ maximum and the following minimum where
it is practically constant, similar to the bulk. 
We note that the $\beta$ 
values are always higher than the relative $b$ values, this is a 
consequence of the fact that the KWW can describe the late $\alpha$  
but not the late $\beta$, early $\alpha$ 
regime where the data are well fitted by the von Schweidler law.

In our correlators oscillations, possibly due to finite size effects, 
in the time zone where the power law of Eq.\ref{eq:gta} holds, prevents us
from calculating the exponent $a$ of MCT directly.
It is still possible to evaluate the parameter 
$a$ from the closure relation of Eq.~\ref{eq:gamma}. 
Using the $\gamma$ calculated from $\tau$ 
and the von Schweidler exponent $b$ found previously,
the values of the critical exponent $a$ for the two species are respectively
$$a_A=0.335, \qquad a_B=0.342$$
These values are quite high, although within the topmost value
indicated by the theory $a\le 0.395$.

\section{Summary and Conclusions}

We have studied the dynamics of a model glass former in confinement.
This study is intended to assess the 
validity of MCT for supercooled liquids embedded in strongly repulsive
and highly confining environments. 
Our model consists in 
a LJBM embedded in a disordered matrix  of soft spheres. 
It represents a possible
modellization for liquids hosted in Silica Xerogels.

The analysis performed in $r$ space and previously reported~\cite{noi} 
showed that the MSD  behavior can be interpreted for 
both A and B particles in terms of
the onset of the cage effect and that the
diffusion coefficients $D$ extracted from the MSD can be fitted
with the power law predicted by MCT. There is however a clear
evidence that the range of validity
of the MCT predictions in the late $\alpha$ relaxation region
suffers a reduction of $60\%$ with respect to the bulk. 
In terms of the small parameter of the theory 
$\epsilon$ the bulk range is 
$0.07<\epsilon<1.30$ while in the confined system it results to be
$0.153<\epsilon<0.631$.
We inferred that the upper bound reduction might be due
to the fact that
B particles tend to avoid the soft sphere interfaces in lowering temperature.
The MCT power law starts in fact to hold for $D$ 
only when most B particles are avoiding 
the soft spheres~\cite{noi}. The lower bound reduction is to be connected to
the presence of important hopping processes 
more marked with respect to the bulk, which tend to introduce deviations
from the ideal MCT predictions earlier than in bulk. 

We have presented here an analysis in $(Q,t)$ space of the confined LJBM
by considering the SISF and we have tested the main scaling relations of MCT.

The static structure factor is similar to that of the bulk phase.
The SISF is well fitted with the stretched exponential function and
the power law behavior of the $\alpha$-relaxation times $\tau$ extracted 
for both A and B particles is consistent with MCT.
Power law fit to both $D$ and $\tau$ as a function of temperature
give the same value for the crossover temperature $T_C=0.355$.
We observe a reduction of circa $20\%$ of $T_C$ in going from bulk
to confined which can be ascribed 
to the existence of an upper bound in confinement for the size of the
domains of cooperative dynamics.
A discrepancy  observed in the values of the exponents $\gamma$
obtained for $D$ and $\tau$
is to be connected to the more important presence of hopping processes in
confinement. Hopping processes influence in fact the behavior of $D$ 
more than that of $\tau$ causing a decrease of the value of the exponent.
This discrepancy was present also in the bulk.

The existence of a confining  
disordered host structure seems to be connected also
to the lowering of the stretching parameter 
possibly caused by a 
larger distribution of relaxation times 
with respect to the bulk.

The behavior of the SISF as a function of $Q$ is very similar to that
of the bulk. All parameters show a monotonic behavior as a function of $Q$ and 
the stretching exponent approaches $b$ for high $Q$.
The exponent $\gamma$ shows no $Q$ dependence.

The SISF plotted against $t/\tau$ shows a scaling 
behavior. The mastercurve can be 
fitted with the von Schweidler function in the late $\beta$, early $\alpha$
region. Analogous to the bulk,
the exponent $b$ is practically constant between the
first maximum and the following minimum of the structure factor.
Similar to  $\beta$ a
substantial reduction of $b$ is observed in going from bulk to confined
LJBM.

The exponent parameter $\lambda$ of the system can be calculated from 
Eq.~\ref{eq:lambda}.
Contrary to MCT prediction we get different values of $\lambda$
on using the exponent $a$ or the exponent $b$ in  Eq.~\ref{eq:lambda}
$$\lambda_{a,A}=0.680, \qquad \lambda_{b,A}=0.871$$
$$\lambda_{a,B}=0.663, \qquad \lambda_{b,B}=0.873$$
This inconsistency is present also in the bulk but it is much less
severe. 
The value of  $\lambda$ reported for the bulk is $0.79\pm0.04$.
Here we have a much larger uncertainty that leads to
$0.8\pm0.1$.

Considering the strong confinement 
experienced by the LJBM, the agreement between MCT predictions 
and the behavior of our confined system as a function of $T$ and $Q$
is most remarkable. 
Differences in behavior with respect to the bulk
on approaching $T_C$ are in fact to be ascribed to the 
presence of more relevant hopping effects.

The possibility to connect the decrease of $T_C$ 
and the enhancement of hopping effects in confinement with
a key role played by dynamical heterogeneities 
in the glass transition scenario is an issue that
would deserve to be addressed in future studies.

\begin{table}
\caption{Values of the parameters of  
Eq.~\protect\ref{eq:tau} extracted from the fit to
$\tau$ and $D$. Values for the bulk system are from Refs.~\cite{kob1,kob2}.
 \label{tab:1}}
 \begin{tabular}{c|cccc}
               &  $T_C^A$  & $T_C^B$ & $\gamma^A$ & $\gamma^B$  \\
 \hline
  Confined system &&&& \\
 \hline
  From $\tau$  &  $0.355$  & $0.355$ & $2.90$      & $2.89$        \\
  From $D$     &  $0.356$  & $0.356$ & $1.86$     & $1.89$       \\
 \hline
  Bulk system &&&& \\
 \hline
  From $\tau$  &  $0.432$  & $0.432$ & $2.6$   & $2.6$        \\
  From $D$     &  $0.435$  & $0.435$ & $2.0$   & $1.7$        \\

 \end{tabular}
 \end{table}


\begin{thebibliography}{}

\bibitem{pablo} For a review on metastable liquids see:
  {P.G. Debenedetti,}
  {\it Metastable Liquids: Concepts and Principles.}
  {Princeton University Press, Princeton }
  {(1997)}.

\bibitem{grenoble}
  See for example the
  {\it Proceedings of the
  International Workshop on ``Dynamics in Confinement''},
  {Eds. B. Frick, R. Zorn, H. B\"uttner,}
  {\it J. Phys. IV} {\bf 10} {(2000)};
  {\it Europhys. J.}, in press {(2003)}.

\bibitem{gotze1} 
{W. G\"otze and L. Sj\"ogren}
{\it Rep. Prog. Phys.}{ \bf 55}, {241} (1992).

\bibitem{gotze2} 
{W. G\"otze} in
{\it Liquids, Freezing and Glass Transition}
{Eds. J. P. Hansen, D. Levesque, J. Zinn-Justin,}
{North Holland, Amsterdam} {(1991)}.

\bibitem{gotze3}
{W. G\"otze,} {\it J. Phys.: Condens. Matter} {\bf 11}, {A1} (1999).

\bibitem{adam-gibbs} G. Adam and J. Gibbs,
{\it J. Chem. Phys.} {\bf 43}, 139 (1965).

\bibitem{donati1} C. Donati, J.F. Douglas, W. Kob, S.J. Plimpton,
P.H. Poole and S.C. Glotzer,
{\it Phys. Rev. Lett.} {\bf 80}, 2338 (1998). 

\bibitem{donati2} C. Donati, S.C. Glotzer and P.H. Poole,
{\it Phys. Rev. Lett.} {\bf 82}, 5064 (1999). 

\bibitem{doliwa1} B. Doliwa and A. Heuer,
{\it Phys. Rev. Lett.} {\bf 80}, 4915 (1998). 

\bibitem{doliwa2} B. Doliwa and A. Heuer,
{\it Phys. Rev. E} {\bf 61}, 6898 (2000). 

\bibitem{berthier} L. Berthier, 
{\it Phys. Rev. Lett.} {\bf 91}, 
055701 (2003). 

\bibitem{gallo1} P. Gallo, E. Spohr and M. Rovere,
{\it Phys. Rev. Lett.}{\bf 85}, 4317 (2000).

\bibitem{gallo2} P. Gallo, E. Spohr and M. Rovere,
{\it J. Chem. Phys.}{\bf 113}, 11324 (2000).

\bibitem{baschnagel}
F. Varnik, J. Baschnagel and K. Binder,
{\it Phys. Rev. E}{\bf 65}, 021507 (2002).

\bibitem{kobconf1}
P. Scheidler, W. Kob and K. Binder,
{\it Europhys. Lett.} {\bf 52}, 277 (2000).

\bibitem{kobconf2}
P. Scheidler, W. Kob and K. Binder,
submitted (2003). cond-mat/0309025.

\bibitem{rosinberg}
{M.L. Rosinberg} in
{\it New Approaches to problems in liquid state theory.}
{Eds. C. Caccamo, J. P. Hansen and G. Stell,}
{Kluwer Academic Publ.}
{(1999)}.

\bibitem{monson}
{L. Sartisov and P. A. Monson}
{\it Phys. Rev. E} {\bf 61}, {7231} (2000).

\bibitem{page}
{K. S. Page and P. A. Monson }
{\it Phys. Rev. E}{ \bf 54}, {R29} (1996).

\bibitem{kob}
{W. Kob and H. C. Andersen},
{\it Phys. Rev. Lett.} {\bf 73}, {1376} (1994);

\bibitem{kob1}
{W. Kob and H. C. Andersen},
{\it Phys. Rev. E} {\bf 51},{4134} (1995);

\bibitem{kob2}
{W. Kob and H. C. Andersen},
{\it Phys. Rev. E} {\bf 52},{4626} (1995).

\bibitem{euro} P. Gallo, R. Pellarin and M. Rovere,
{\it Europhys. Lett.} {\bf 57}, 212 (2002).

\bibitem{noi}
P. Gallo, R. Pellarin and M. Rovere,
{\it Phys. Rev. E} {\bf 67} 041202 (2003).


\bibitem{fucs} M. Fucs,
{\it J. Non-Cryst. Solids} {\bf 172-174}, 241 (1994).

\end{thebibliography}
\end{document}